# Comparison of Feature Point Extraction Algorithms for Vision Based Autonomous Aerial Refueling

[1]Borui Li, [2]Chundi Mu, [3]Tao Wang, [4]Qian Peng

[1, First Author]Department of Automation, Tsinghua University, 100084, Navy Academy of Armament, 100036, China, lbr07@mails.tsinghua.edu.cn

[*2,Corresponding Author]Department of Automation, Tsinghua University, 100084, China, muchd@mail.tsinghua.edu.cn

[3,4] Navy Academy of Armament, 100036, China

*Abstract*
*Autonomous aerial refueling (AAR) is the most effective way to extend unmanned aerial vehicles' (UAV) endurance and improve their mission performance. A monocular vision based relative pose estimation (MVRP) method is proposed to solve the problem of close-range and high-accuracy relative pose estimation, which is very important and critical for AAR. MVRP is a feature point tracking based method, which makes feature point extraction algorithm become a key issue for MVRP method. An MVRP virtual reality simulation system is set up based on Vega Prime. This paper carries out several simulation experiments to compare three feature point extraction algorithms within the MVRP method. The simulation results show that FAST algorithm is able to get a good balance between speed and accuracy, compared with the other two algorithms.*

## 1. Introduction

In recent years, the technology of unmanned aerial vehicle (UAV) has developed rapidly in both military and civil areas. UAVs' major drawbacks are payload shortage and short endurance. Autonomous aerial refueling (AAR) is the most effective way to overcome these drawbacks [1, 2]. There are mainly two aerial refueling approaches: the probe-and-drogue refueling (PDR) approach and the boom-and-receptacle refueling (BRR) approach. Compared with PDR, BRR has faster refueling speed, higher success rate and is more insensitive to bad weather and atmospheric turbulence, which makes it the trend and focus of AAR research. The process of UAV-AAR using the BRR method could be divided into four phases: rendezvousing, docking, refueling and dismissing [3]. In the docking phase, the receiver aircraft (UAV) move from the "pre-contact position" behind and below the tanker to the "contact position" which is much closer to the tanker [4]. In the refueling phase, the UAV has to stabilize within a small enough neighborhood region of the contact position. The distance between the tanker and UAV are pretty close in docking and refueling phase. Therefore, it makes critical and high demands on accuracy of relative pose estimation between two aircraft [2, 5], which is a key issue of UAV-AAR. Traditional navigation systems such as GPS (global position system) and INS (inertial navigation system) is not suitable to AAR because their accuracy is low and the GPS signal is easy to be blocked by the tanker or jammed. Vision based technology provides an effective way to overcome traditional navigation systems' shortcomings and improve the accuracy of relative pose estimation.

Vision based relative pose estimation methods for UAV-AAR can be divided into two categories: image based method and feature tracking based method. Image based method fuses the characteristics of the image itself, such as gray values, with data from other sensors [6]. It doesn't require model parameters, but it's quite difficult to implement. In UAV-AAR, artificial identification markers could be mounted on both the tanker and the UAV to help the process of feature extracting and tracking. Feature tracking based method is easier to understand and implement than image based method. Its basic idea is to estimate the relative pose information by extracting and tracking specific features in images, such as dots, lines and so on. Camera parameters and aircraft models are required for utilizing feature tracking based method. Doebbler et al. [7] use the active deformable contour algorithm to extract a rectangular marker painted on the nose of the UAV, and track the marker's center of mass to estimate the relative pose information. However, this method is only capable for refueling phase and can't provide all the six degrees of freedom information. Campa et al. [4, 8, 9] install some passive optical markers on the tanker. After detecting and tracking these markers by some machine vision algorithms, relative pose information is calculated using the iterative algorithms such as Gauss-Newton and OI algorithm [10]. The simulation results show that centimeter level tracking error can be achieved. However, this method doesn't consider eliminating the gross errors that may occur in the process of matching the markers. Thus, if some markers are missed or mismatched, the accuracy of relative pose information will be significantly decreased or even unavailable. Ding et al. [11] implement and test OI algorithm within a simulation environment for relative pose estimation of UAV-AAR. The impact of feature points' configuration and number on pose estimation accuracy is discussed. But the simulation results are not very credible. They didn't mention how to obtain the coordinates of the feature points from the images. Most of the configurations are meaningless because nearly all of the feature points will move out of the camera's field of view when the two aircraft are getting close in docking phase.

To solve the close-range relative pose estimation problem of UAV-AAR, a monocular vision based relative pose estimation (MVRP) method is presented in this paper. MVRP is a feature point tracking based method. It selects some navigation signal beacons on the tanker as artificial identification markers, which are called pre-set feature points (PFPs). The relative pose information between the tanker and the UAV is estimated by extracting and tracking the PFPs. Feature extraction algorithm is the key issue in MVRP method and observably impact on the performance of MVRP method. A three-dimensional (3D) visual simulation system based on virtual reality software Vega Prime is set up to test the MVRP method. Three commonly used feature point algorithms are compared within the MVRP simulation system.

## 2. Vision-based AAR relative pose estimation problem

### 2.1. Reference frames

All the coordinate values involved in this paper will be expressed in homogeneous coordinates. A set of reference frames are need to be defined to describe the AAR relative pose estimation problem, as shown in Figure 1.

(1) Tanker reference coordinate frame $O_T X_T Y_T Z_T$. The origin $O_T$ is located at the tanker's center of mass, while the $X_T$ axis points right along the tanker's horizontal axis, $Y_T$ axis points left along the longitudinal axis and $Z_T$ axis points up along horizontal axis.

(2) The definition of UAV reference coordinate frame $O_U X_U Y_U Z_U$ is similar to $O_T X_T Y_T Z_T$.

(3) Camera reference coordinate frame $O_C X_C Y_C Z_C$. The camera is mounted on the UAV with a pitch angle of $\alpha$ to capture images of the tanker. To make further discussion simplified, we assume that $O_C$ and $O_U$ are overlapped.

(4) The image plane coordinate frame $O_I X_I Y_I$ is centered at the top left corner of the image. The $X_I$ axis points right and is parallel to $X_C$ axis, while The $Y_I$ axis points down and is parallel to $Y_C$ axis.

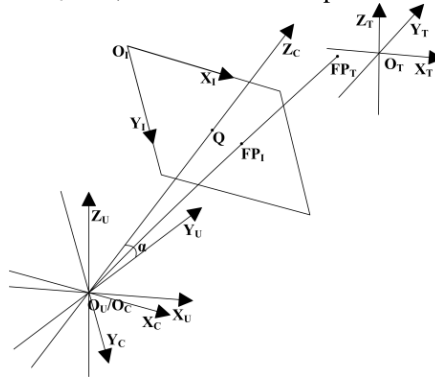

**Figure 1.** Reference frames used for AAR relative pose estimation

## 2.2. Problem description

Let $X(k)$ denote the tanker's pose information in UAV reference coordinate frame at sample time $k$:
$$X(k) = (x_k, y_k, z_k, \psi_k, \theta_k, \varphi_k)^T,$$
where $[x_k, y_k, z_k]^T$ refers to the 3D position coordinates and $[\psi_k, \theta_k, \varphi_k]^T$ refers to the attitude angles.

Similarly, let $CP_U = (x_U, y_U, z_U, \psi_U, \theta_U, \varphi_U)^T$ denote the camera's pose information in the UAV reference coordinate frame. Some navigation signal beacons installed on the underside of the tanker are selected as the PFPs, and the number of them is $N$. Let $FP_T^i = (x_T^i, y_T^i, z_T^i, 1)^T$, $i = 1..N$ denote the PFPs' 3D position coordinates in the tanker reference coordinate frame.

Thus, the relative pose estimation problem in UAV-AAR could be summarized as follows:
Given $CP_U$ and $FP_T^i, i = 1..N$, solve $X(k)$.

## 3. Feature point extraction algorithm

The MVRP method selects navigation signal beacons as PFPs and the PFPs on images fit the characteristics of "corner", so generic corner extraction algorithms can be used to extract PFPs from captured images. Three typical and frequently-used corner extraction algorithms are introduced in this section. Since some undesired corners will be extracted and some PFPs may be missed, the feature point set derived by generic corner extraction algorithms contains all or part of PFPs.

The captured images will be converted from RGB color space to HSV color space at first, and then the V channel data is picked out to utilize the feature point extraction algorithms.

### 3.1. Harris algorithm

Harris algorithm is proposed by Harris and Stephens [12]. Let $I_U$ and $I_V$ denote the first gradients of image intensity along the $X_I$ and $Y_I$ axis of image plane respectively. Harris algorithm [13] takes a small window to each pixel $(u,v)$ and calculates the image gray gradient matrix $M_H$:
$$M_H = \sum_{u,v} w(u,v) \begin{bmatrix} I_U^2 & I_U I_V \\ I_U I_V & I_V^2 \end{bmatrix}.$$
where $w(u,v)$ is a Gaussian window function.

Then the corner response function of $(u,v)$ can be expressed as:
$$R(u,v) = \det(M_H) - k_H * [Tr(M_H)]^2,$$
where $\det(M_H)$ refers to the determinant of $M_H$, $Tr(M_H)$ refers to the trace of $M_H$, and $k_H$ is a specific small parameter which could be set to 0.04~0.06 based on experience.

If $R(u,v)$ is greater than a certain threshold, then the pixel $(u,v)$ is a corner.

### 3.2. SUSAN algorithm

Smith and Brady [14] proposed a novel corner detection algorithm called smallest univalve segment assimilating nucleus (SUSAN) algorithm [15]. Considering a 37-pixel or 25-pixel circular mask with a center pixel called the "nucleus", the area of pixels which have the same (or similar) brightness as the nucleus is defined as "univalve segment assimilating nucleus (USAN)". The basic idea of SUSAN algorithm is to find the pixels with small enough USAN.

Let the circular mask slide on each pixel to calculate the comparison function:
$$c(r, r_0) = e^{-(\frac{I(r)-I(r_0)}{t})^6},$$
where $r_0$ denotes the nucleus pixel, r denotes any other pixel within the circular mask, $I(r)$ and $I(r_0)$ denote the gray value of $r$ and $r_0$. The USAN size of $r_0$ can be described as follows:

$$n(r_0) = \sum_r c(r, r_0).$$

Then the corner response of $r_0$ is calculated by:

$$R(r_0) = \begin{cases} g - n(r_0), & \text{if } n(r_0) < g \\ 0, & \text{if } n(r_0) \geq g \end{cases},$$

where $g$ is called the geometric threshold and usually set to $\frac{1}{2}\max(n(r_0))$, which means $g=18.5$ when using the 37-pixel mask and $g=12.5$ when using the 25-pixel mask.

At last, SUSAN algorithm performs non-maximum suppression using a 5*5 pixel template to find local positive maxima, i.e. corners.

### 3.3. FAST algorithm

Rosten and Drummond [16] proposed a simple and very fast corner detection algorithm called FAST. FAST algorithm defines "corners" as the pixel with a sufficient number of pixels in its neighborhood whose image intensity is greater enough or less enough than the candidate pixel by some threshold.

For each pixel $P$, consider the 16 pixels on the circle around $P$ with a radius of 3 pixels, as illustrated in Figure 2. Denoting the image intensity of $P$ by $I(P)$, calculate $N_{F,B}(P)$ and $N_{F,D}(P)$ with the following formula:

$$N_{F,B}(P) = \sum_{\forall x \in circle(P,3)} Sgn(I(x) - I(P) > \varepsilon)$$
$$N_{F,D}(P) = \sum_{\forall x \in circle(P,3)} Sgn(I(P) - I(x) > \varepsilon),$$

where $\varepsilon$ is the intensity difference threshold, which is used to determine whether the intensity difference between pixel $x$ and $P$ is large enough or not. The function $Sgn(*)$ is defined as follows: if the event in the parentheses is true, then $Sgn(*)=1$; else $Sgn(*)=0$.

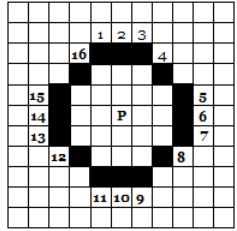

**Figure 2.** The schematic diagram of FAST algorithm

If $N_{F,B}(P)$ (or $N_{F,D}(P)$) is greater than a certain threshold $t_F>0$ and at least $t_F$ pixels among these $N_{F,B}(P)$ (or $N_{F,D}(P)$) pixels are contiguous, then $P$ is a corner. $t_F$ is usually set to 3/4 of the number of pixels on the circle around $P$

## 4. MVRP method

The flow chart of MVRP method is shown in Figure 3. At sample time $k$, the MVRP method is as follows:

Step 1: Calculate the current camera imaging model according to current estimated value of $X(k)$.

Step 2: Calculate the PFPs' coordinates in image plane coordinate frame according to the current camera imaging model and $FP_T^i, i=1..N$.

Step 3: Extract the PFPs from the images of the tanker, and obtain their coordinates in image plane coordinate frame. This step is called FE for short.

Step 4: Match the two set of coordinates respectively derived in Step 2 and Step 3, and eliminate gross matching errors. This step is called MGE for short.

Step 5: Estimate $X(k)$ by iterative pose estimation algorithm. If the termination conditions of iterative pose estimation algorithm are met, stop the iteration and output the relative pose information $\hat{X}(k)$; else return to Step 1.

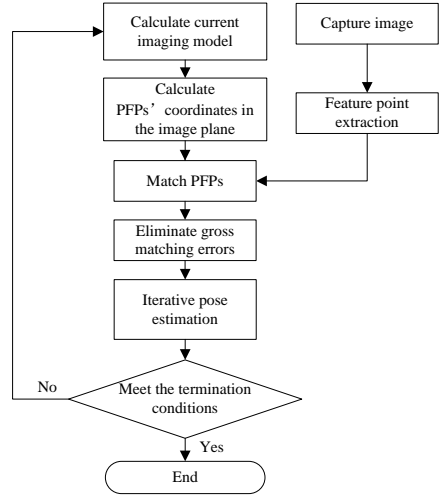

**Figure 3.** The flow chart of the MVRP method

### 4.1. Camera imaging model

The virtual camera in Vega Prime could be modeled by the standard pin-hole model [17], as shown in Figure 1. The intersection of $O_C Y_C$ and the image plane $O_I X_I Y_I$ is called the principal point, denoted by $Q = (u_0, v_0, 1)^T$.

For each PFP $FP_i$, its coordinates could be expressed in tanker coordinate frame and camera coordinate frame as follows:
$$FP_T^i = (x_T^i, y_T^i, z_T^i, 1)^T,$$
$$FP_C^i = (x_C^i, y_C^i, z_C^i, 1)^T.$$

Let $FP_I^i = (u_I^i, v_I^i, 1)^T$ denote the coordinates of $FP_i$'s image point in image plane coordinate frame. According to the principle of similar triangles, we can easily have:

$$FP_I^i = \frac{1}{z_C^i} \begin{bmatrix} f_x & 0 & u_0 & 0 \\ 0 & f_y & v_0 & 0 \\ 0 & 0 & 1 & 0 \end{bmatrix} FP_C^i.$$

Let $K$ denote $\begin{bmatrix} f_x & 0 & u_0 & 0 \\ 0 & f_y & v_0 & 0 \\ 0 & 0 & 1 & 0 \end{bmatrix}$, then $K$ is called the intrinsic matrix of the virtual camera and we have:

$$FP_I^i = \frac{1}{z_C^i} K * FP_C^i.$$

Let $M$ denote the transformation matrix between $FP_T^i$ and $FP_C^i$, then $M$ is the extrinsic parameter matrix of the virtual camera and we have:

$$FP_C^i = M * FP_T^i.$$

Thus,

$$FP_I^i = \frac{1}{z_C^i} K * M * FP_T^i.$$

Let $M_r(\beta, S)$ refer to the rotation matrix of rotating angle $\beta$ around the S-axis counter-clockwise, seeing from the S-axis' positive direction. Then the formula for calculating $M$ according to $X(k)$ and $CP_U$ is as follows:

$$M = M_e * M_r(\alpha, X_U) * M_t * M_r(\psi_k, Z_T) * M_r(\theta_k, X_T) * M_r(\varphi_k, Y_T),$$

where $M_e = \begin{bmatrix} 1 & 0 & 0 & 0 \\ 0 & 0 & -1 & 0 \\ 0 & 1 & 0 & 0 \\ 0 & 0 & 0 & 1 \end{bmatrix}$ and $M_t = \begin{bmatrix} 1 & 0 & 0 & x_k \\ 0 & 1 & 0 & y_k \\ 0 & 0 & 1 & z_k \\ 0 & 0 & 0 & 1 \end{bmatrix}$.

Then we can have a 2*N matrix $P_P$:

$$P_P = \left[ FP_I^1, FP_I^2, \cdots, FP_I^N \right].$$

It's easy to know that $M$ and $P_P$ are functions of $X(k)$.

### 4.2. PFP matching and gross error eliminating

When the 2D coordinates of projected feature points $P_P$ are derived, the next step would be matching them to the extracted feature points gained by a feature point extraction algorithm, denoted by $P_E$. Since there is a one-to-one correlation between the projected feature points and the physical PFPs on the tanker, this step is equivalent to matching the extracted feature points to the physical PFPs on the tanker. If a projected feature point $p_i$ and an extracted feature point $e_j$, which are also column vectors of $P_P$ and $P_E$, are corresponding feature points, they should satisfy the following conditions simultaneously [8]:

(1) $p_i$ is the nearest projected feature point to $e_j$ among all the projected feature points (column vectors) of $P_P$.

(2) $e_j$ is the nearest extracted feature point to $p_i$ among all the extracted feature points (column vectors) of $P_E$.

Rearrange the matched feature points' coordinates in the same order, then we get matrixes $P_P^{'}$ and $P_E^{'}$. There may be gross matching errors in $P_P^{'}$ and $P_E^{'}$, for example: if a PFP is not extracted by the feature point extraction algorithm, its projected feature point still might be matched to another extracted point, which satisfies the matching conditions but is not the desired "corresponding extracted point".

Since the camera imaging model is calculated through the estimated value not the real value of $X(k)$, and the pin-hole model is an approximate imaging model, the Euclidean distance between a pair of matched points is usually not zero. However, the variance of these distances wouldn't be large. Based on these characteristics, the following algorithm is used to eliminate the gross matching errors:

Step 1: Calculate the Euclidean distance between each pair of matched points, denoted by $d_i$.

Step 2: For each pair of matched points, calculate the average distance $\bar{d}$ of the other matched point pairs, if $d_i > T_1$ and $\frac{d_i - \bar{d}}{\bar{d}} * 100\% > T_2$, then this pair of matched points is a gross error. $T_1$ and $T_2$ are pre-set thresholds.

Step 3: If gross error exists, remove the wrong matched point pairs from $P_P^{'}$ and $P_E^{'}$, and return to Step 2; else stop the iteration.

### 4.3. Iterative pose estimation

Let $N_M$ denote the number of right matched feature point pairs after eliminating gross errors. Rearrange the coordinates in $P_P^{'}$ and $P_E^{'}$ in the same order, then we get two $2N_M*1$ column vectors, still denoted by $P_P'$ and $P_E'$. $P_P$ is a function of $X(k)$, so $P_P'$ is also a function of $X(k)$, then the relative pose estimation problem of UAV-AAR could be described as the following optimization problem:

$$\arg\min_{X(k)} \left\| \frac{P_E' - P_P'(X(k))}{N_M} \right\|^2.$$

This optimization problem could be solved by iterative algorithms such as Newton-Gauss, Levenberg-Marquard, OI algorithm and so on. MVRP method choose Levenberg-Marquard (L-M) algorithm to estimate the pose information. It should be noted that L-M algorithm requires that the dimension of $P_E' - P_P'(X(k))$ is greater than $X(k)$, so relative pose information could be estimated only when $N_M \geq 3$.

L-M algorithm could converge to sufficiently accurate result as long as the initial value of $X(k)$ is chosen properly. The initial value of $X(k)$ could be chosen with the following instructions:

(1) If $k=1$, since the distance between the tanker and UAV is still far away enough, the data of GPS/INS would be available, and they could be used as the initial value.

(2) If $k \geq 2$, the final estimation of relative pose information at previous sample time $k$-1 could be used as the initial value, because the relative pose information between two consecutive sample times would not change significantly because of the high-frequency updates of MVRP method.

## 5. Simulation experiments

In order to compare the performance of the three feature point extraction algorithms (Harris, SUSAN and FAST) in MVRP method, we set up a virtual reality simulation system based on Vega Prime and carry out a series of simulation experiments. Vega Prime is an outstanding real-time 3D development software suite, which could be easily configured and developed using Visual C++. A 3D KC-10 aircraft model is employed as the tanker model, and an F-16 fighter model is rescaled to match the size of the UAV. The algorithms of feature point extracting, matching and gross error eliminating are coded by Visual C++. The Vega Prime configuration and virtual camera simulation program are realized by Visual C++, while the L-M algorithm is implemented by MATLAB. A graphic workstation with 2.21GHz quad-core CPU, 2GB RAM and Quadro FX5600 graphics card is used for simulation.

### 5.1. Simulation parameters

7 navigation signal beacons which are fixed on the underside of the tanker are used as the pre-set feature points, and their 3D coordinates in tanker reference coordinate frame are presented in Table 1.

**Table 1.** 3D coordinates in tanker reference coordinate frame of PFPs

| $i$ | $x_T^i$ /m | $y_T^i$ /m | $z_T^i$ /m |
|---|---|---|---|
| 1 | 0 | 15.79 | -1.83 |
| 2 | 2.69 | 6.75 | 0.11 |
| 3 | -2.69 | 6.75 | 0.11 |
| 4 | 16.12 | -6.98 | -0.94 |
| 5 | 6.2 | -4.33 | -2.48 |
| 6 | -6.2 | -4.33 | -2.48 |
| 7 | -16.12 | -6.98 | -0.94 |

The virtual camera is mounted on the nose of the UAV with a pitch angle of +38°, looking up to capture images of the tanker. The camera's intrinsic parameter matrix can be derived according to its relationship with the projection matrix $R$ of Vega Prime. Let $W$ denote the width of the image, $H$ denote the height of the image, and assign the principal point coordinates $(u_0, v_0)$ to $(W/2, H/2)$, then we have $f_x = (R_{1,1} * W)/2$ and $f_y = (R_{2,2} * H)/2$.

Assign the relative pose information at the pre-contact position to (0,80,60,0,0,0) and the contact position to (0,40,20,0,0,0). The measurement unit is meter for relative position components, and degree for relative attitude angles. From (0,80,60) to (0,45,25), capture a image when the coordinates' change reaches (-0.5,-0.5,0). 71 images are captured in all and the image size is 512*384.

The thresholds used for gross error eliminating are set to $T_1$=5 pixels and $T_2$=50%.

### 5.2. Speed performance

Apply MVRP method to each image using the three feature point extraction algorithms respectively. The average time required by Step FE, denoted by $t_{FE}$, is shown in Table 2. In general, Harris algorithm required much more time than the others, while the time required by SUSAN algorithm is nearly the same as FAST algorithm.

**Table 2.** Average time required by feature point extraction algorithms

|  | **Harris** | **SUSAN** | **FAST** |
|---|---|---|---|
| $t_{FE}$ /s | 0.0934 | 0.0453 | 0.0455 |

### 5.3. Accuracy

The accuracies of the three feature point extraction algorithms are compared in two aspects: the number of PFPs which are not extracted and the relative pose estimation accuracy. The accuracy comparison results are shown in Table 3, where $N_{miss}$ denotes the average number of PFPs which are not extracted by feature point extraction algorithms and $|e_x| \sim |e_\varphi|$ denote the average absolute errors of six components of pose information.

**Table 3.** Accuracy performance of feature point extraction algorithms

|  | $N_{miss}$ | $|e_x|$ /m | $|e_y|$ /m | $|e_z|$ /m | $|e_\psi|$ /° | $|e_\theta|$ /° | $|e_\varphi|$ /° |
|---|---|---|---|---|---|---|---|
| Harris | 0.90 | 0.01 | 0.25 | 0.15 | 0.15 | 0.41 | 0.14 |
| SUSAN | 0.03 | 0.04 | 0.49 | 0.29 | 0.23 | 0.53 | 0.26 |
| FAST | 1.72 | 0.03 | 0.32 | 0.17 | 0.30 | 0.90 | 0.24 |

The error curves of relative pose estimation are shown in Figure 4~Figure 9. In general, the accuracy of Harris algorithm is the best, and the average error of MVRP method using SUSAN algorithm is roughly greater than the other two algorithms. When the feature point extraction algorithm failed to extract all the visible PFPs, the MVRP method could still work normally. The figures demonstrate that no matter which feature point extraction algorithm is used, the estimation errors of position components could achieve the required accuracy and the errors of attitude angles don't exceed 1°. It means that the MVRP method can meet the accuracy demand of AAR.

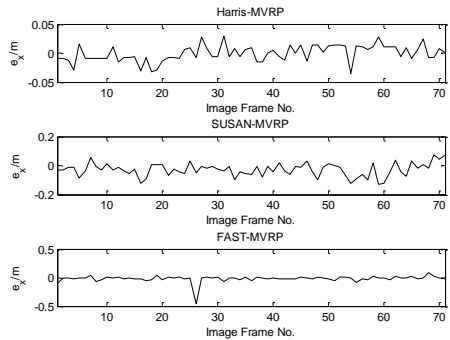

**Figure 4.** Estimation error comparison of position component *x*

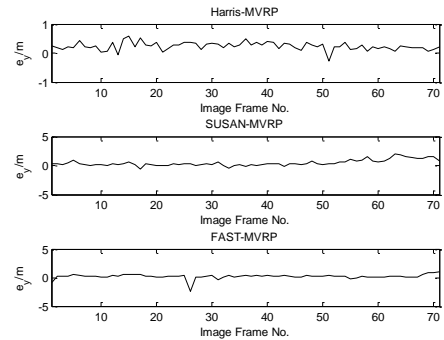

**Figure 5.** Estimation error comparison of position component y

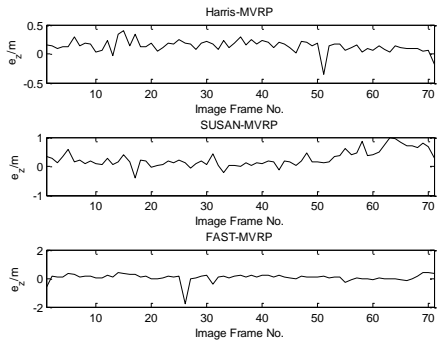

**Figure 6.** Estimation error comparison of position component *z*

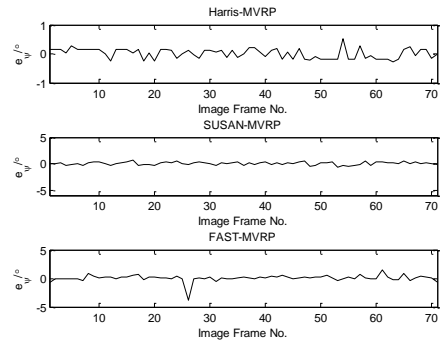

**Figure 7.** Estimation error comparison of heading angle $\psi$

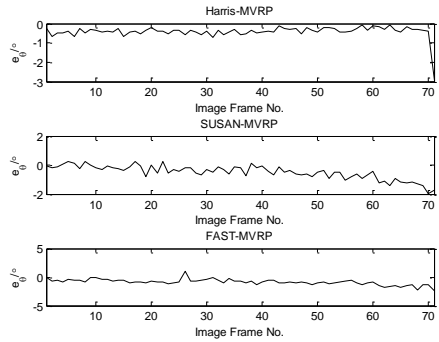

**Figure 8.** Estimation error comparison of pitch angle $\theta$

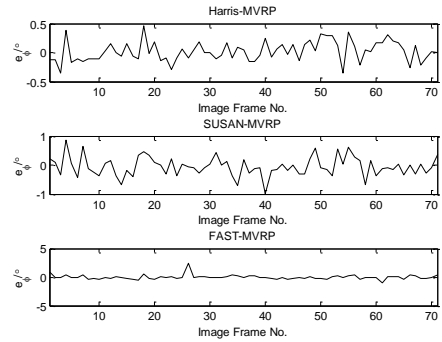

**Figure 9.** Estimation error comparison of roll angle $\varphi$

## 6. Conclusions

This paper compares three frequently-used feature point extraction algorithms within the MVRP method. MVRP is a feature point tracking based relative pose information estimation method for UAV-AAR, and feature point extraction algorithm is a quite important component of MVRP method. According to the simulation experiment results, we can draw the following conclusions:

(1) No matter which one of the three frequently-used feature point extraction algorithms is utilized, the MVRP method is able to meet the accuracy requirement of UAV-AAR. The MVRP method is easy to implement, and doesn't need to change the aircraft body. However, the MVRP method would not work well when two aircraft are far away from each other (e.g. more than 200 m), and its performance is sensitive to the initial value of $X(k)$.

(2) Compared with Harris and SUSAN algorithm, FAST algorithm can provide a good tradeoff between speed and accuracy performance.

(3) The MVRP method can still work normally when some of PFPs are mismatched, or not extracted, or leave the camera's field of view, as long as the number of right-matched PFPs is greater than or equal to 3.